\documentclass[5p,times]{elsarticle}
\usepackage{lineno,hyperref}

\usepackage{multicol}
\usepackage{multirow}
\usepackage{array}
\usepackage{ulem}
\usepackage[dvipsnames]{xcolor}
\usepackage{mdwmath}
\usepackage{mdwtab}
\usepackage{eqparbox}
\usepackage{url}
\usepackage{algorithmic}
\usepackage{textcomp}
\usepackage[ruled,vlined]{algorithm2e}
\usepackage{caption}
\usepackage{subcaption}
\usepackage{natbib}
\usepackage{colortbl}
\usepackage{url}
\usepackage{float}
\usepackage{booktabs}
\usepackage{amsmath}
\usepackage{mathtools}
\usepackage{float}
\usepackage{mathptmx}
\usepackage{mathrsfs}
\usepackage{comment}
\usepackage{amsmath}
\usepackage{amssymb}
\usepackage{listings}

\journal{Journal of \LaTeX\ Templates}

\begin{document}

\begin{frontmatter}

\title{Autoencoder-based Unsupervised Intrusion Detection using Multi-Scale Convolutional Recurrent Networks} 


\author[mymainaddress]{Amardeep Singh}

\author[mymainaddress]{Julian Jang-Jaccard\corref{mycorrespondingauthor}}
\cortext[mycorrespondingauthor]{Corresponding author}
\ead{j.jang-jaccard@massey.ac.nz}

\address[mymainaddress]{Cybersecurity Lab, Massey University, Auckland, NEW ZEALAND}

\begin{abstract}
The massive growth of network traffic data leads to a large volume of datasets. Labeling these datasets for identifying intrusion attacks is very laborious and error-prone. Furthermore, network traffic data have complex time-varying non-linear relationships. The existing state-of-the-art intrusion detection solutions use a combination of various supervised approaches along with fused features subsets based on correlations in traffic data. These solutions often require high computational cost, manual support in fine-tuning intrusion detection models, and labeling of data that limit real-time processing of network traffic. Unsupervised solutions do reduce computational complexities and manual support for labeling data but current unsupervised solutions do not consider spatio-temporal correlations in traffic data.
To address this, we propose a unified Autoencoder based on combining multi-scale convolutional neural network and long short-term memory (MSCNN-LSTM-AE) for anomaly detection in network traffic. The model first employs Multiscale Convolutional Neural Network Autoencoder (MSCNN-AE) to analyze the spatial features of the dataset, and then latent space features learned from MSCNN-AE employs Long Short-Term Memory (LSTM) based Autoencoder Network to process the temporal features. Our model further employs two Isolation Forest algorithms as error correction mechanisms to detect false positives and false negatives to improve detection accuracy. 
We evaluated our model  NSL-KDD, UNSW-NB15, and CICDDoS2019 dataset and showed our proposed method significantly outperforms the conventional unsupervised methods and other existing studies on the dataset.
\end{abstract}


\end{frontmatter}

\section{Introduction}

An intrusion detection system (IDS) is a primary defense mechanism against any expected or unexpected cyberattacks to adapt and secure the computing infrastructures \cite{8369054, jang2014survey}. IDS based on machine/deep learning is currently the main focus in recent research studies in intrusion detection as they are more effective against large-scale attacks. Various studies based on support vector machine, K-nearest neighbor, decision tree, random forest, and deep learning models such as convolutional neural network-based, a recurrent neural network have been applied for building IDS \cite{electronics9010173,article11,article13}. The Intrusion detection system suffers from complexity issues (a large amount of high dimensional complex data) and is insufficient for learning complex nonlinear relationships that change over time between large datasets \cite{IERACITANO202051}. As Hogo \cite{6987012} stated in his work, there are temporal and spatial dependencies in the traffic data \cite{article4} but existing IDS mostly focus on the last snapshot. Like, source IP and destination IP define the subject and object of the behavior in the data streams, and the duration describes how long the behavior lasted \cite{Rieman1}. Similarly, the packet volume and packet size indicate the traffic flow, and their size varies between different protocols \cite{Rieman1}. Therefore, this information should be analyzed in the context of the communication protocol to explain the impact of this behavior on the communication capability. This spatio-temporal relationship between the multivariate data helps in detecting various attack characteristics that look rather benign under different network protocols \cite{Rieman1}.\\
The existing state-of-the-art methods address this issue by proposing feature fusion and association to improve the ability to identify malicious network behaviors \cite{li2018data}. For example, Li et. al \cite{LI2020107450} proposed IDS that divide features into four subsets based on the correlation among features and used a multi-convolutional neural network to detect intrusion attacks. Similarly, Wei et al. \cite{8171733} proposed fusion of convolutional neural network (CNN) to learn spatial features and long short term memory (LSTM) to learn the temporal features among multiple network packets.
Mostly machine learning (ML) or deep learning (DL) solutions proposed for the IDS are based on a supervised learning approach.
Unfortunately, the IDS models based on supervised ML techniques liaise heavily on manually labeled raw traffic and fine-tuning of ML models based on the labeled datasets. This human-engineered network traffic labeling is very laborious considering the massive growth of traffic data and may lead to error-prone data labels \cite{binbusayyis2021unsupervised}. This stems from the " Unsupervised intrusion detection" approach of trying to define what is anomalous rather than what is normal \cite{5485505}. 
In recent years, with the advent of deep architectures such as fully connected Autoencoders and self-organizing maps, there have been substantial advances in the domain of unsupervised IDS. However, unsupervised models do not consider spatial and temporal characteristics of traffic data.

We address this issue by proposing a new model that utilizes three different deep learning models.
The main contributions of this paper can be summarised as follows.

\begin{itemize}
\item We propose a unified Autoencoder based on combining multi-scale convolutional neural network and long short-term memory (MSCNN-LSTM-AE)  for anomaly detection in network traffic. Our proposed model runs in an unsupervised manner by effectively removing the requirement of having to manually label the data.
\item The MSCNN-AE part of our model is capable of extracting inherited spatial patterns from network traffic. The LSTM-AE part of our model extracts the temporal patterns from network traffic in addition to the spatial patterns obtained through the MSCNN-AE. We utilize Autoencoder for both MSCNN and LSTM training without labels (i.e. unsupervised).

    \item To further improve classification accuracy, we utilize a two-staged detection technique using isolation forest to effectively reduce the false positives and false negatives created by the threshold-based approach used in the Autoencoder.

    \item Our experimental results, benchmarked on NSL-KDD, UNSW-NB15, and CICDDoS2019 datasets, show that our proposed model achieves a higher recall and f-score compared to other state-of-the-art models.

\end{itemize}

The rest of the paper is organized as follows. Section $II$ presents related works for intrusion detection. Section $III$ gives a brief description of the typologies of deep neural networks employed in the proposed approach.  Section $IV$ introduces our proposed model. Section $V$, describes datasets, experimental setup, and evaluation metrics. In section $V$, the experimental results are discussed and compared to existing studies. Lastly, section $VI$ draws the conclusions of our proposed method.

\section{Related work}\label{sec:rw}
The recent literature on cybersecurity reveals how the advancements in unsupervised AI techniques have led the intrusion detection research. For example, Auskalnis et al. \cite{article343} use the Local Outlier Factor during data preprocessing to exclude normal packets that overlap with the density position of anomalous packets. Later, a cleaned (reduced) set of normal packets is used to train another local outlier model to detect anomalous packets. Similarly, Rathore et al. \cite{article314} proposed unsupervised IDS that is based on semi-supervised fuzzy C-Mean clustering with single hidden layer feedforward neural networks (also known as Extreme Learning Machine) to detect intrusions in real-time. Aliakbarisani et al. \cite{article308} proposed a method that learns a transformation matrix based on the Laplacian eigenmap technique to map the features of samples into a new feature space, where samples can be clustered into different classes using data-driven distance metrics.\\
With the advent of deep learning methods and cheap hardware (graphical processing units), unsupervised deep learning methods such as deep belief network (DBN), self-organizing maps, and autoencoders are increasingly being used. Karami \cite{article315} proposed an IDS model based on a self-organizing map that not only removes benign outliers but also improves the detection of anomalous patterns. Alom et al. \cite{7443094} proposed an unsupervised deep belief network (DBN) for intrusion detection. Similarly, for the in-vehicular network security, Kang et al. \cite{kang2016intrusion} leveraged the benefit of the unsupervised pretraining process of the DBN model. Additionally, a considerable number of studies have investigated the application of DBN in IDS design. \cite{gao2014intrusion, zhang2017deep}.
Nevertheless, recent studies have focused primarily on autoencoder (AE) to develop efficient IDS because they are easy to implement and inexpensive computational cost. A number of studies have attempted to develop variants of AE with improved discriminative intrusion detection.
For instance, Hassan et al. \cite{electronics9020259} optimized hyper-parameters of sparse AE to extract better feature embedding for classifying intrusion attacks.
Similarly, Song et al. \cite{article7} proposed an Autoencoder model (trained on normal samples) based on the principle that the reconstruction loss of normal traffic samples is lower than that of abnormal (attack) samples so that a threshold can be set for detecting future attacks. In addition, this work evaluates various hyperparameters, model architectures, and latent size settings in terms of attack detection performance. The various researcher proposed methods that combine unsupervised and supervised methods to get the best of both learning techniques.
Like, Shone et al. \cite{article331}  proposed unsupervised feature learning using a non-symmetric stacked deep autoencoder. Moreover, these features are used for intrusion detection using a random forest model.
Similarly, Hawawreh et al. \cite{article321} combined autoencoder with a deep feed-forward neural network for intrusion detection.
Sheng et al. \cite{mao2019discriminative} developed a framework that combines generative adversarial networks and Autoencoder for improving the performance for intrusion detection.
Aygun et al. \cite{aygun2017network} introduced AE  variant that stochastic-ally determined threshold for reconstruction error to determine intrusion attack and improve the discriminative ability of AE on NSL-KDD intrusion datasets.
Ieracitano et al. \cite{IERACITANO202051} used statistical analysis to select more relevant features as well as filtered out local outliers before training the AE variant. This methodology improved performance on the NSL-KDD dataset.
Mirsky et al. \cite{mirsky2018kitsune} proposed an ensemble of autoencoders to detect intrusion attacks.
In the same vein, Al-Qatf et al. \cite{al2018deep} proposed a combination of AE with support vector machine (SVM) model. They used sparse AE for extracting meaningful feature embedding and used an SVM classifier with encoded features for classifying intrusion attacks. Their proposed combination of AE and SVM can efficiently be used for binary and multi-class scenarios.
Moreover, the authors in \cite{qureshi2020intrusion} applied a sparse AE model exploiting the concepts of self-taught learning to learn useful features for intrusion detection. In addition, they combined the original features with extracted features to improve the model generalization ability in recognizing the network attacks. Furthermore, in \cite{kherlenchimeg2020deep}, a two-stage framework that combines a sparse AE with long short-term memory is investigated for building an efficient IDS. Here, the framework employs the sparse AE for learning effective feature representation and the LSTM model for classifying normal and malicious traffic. In a work by Shuaixin \cite{shuaixin2020intrusion}, the viability of combining stacked AEs and an SVM classifier configured with a piece-wise radial basis function to improve the classification performance of the SVM for intrusion detection is examined. Similarly, the authors in \cite{yu2017network} combined the advantage of stacked AEs with a CNN to considerably achieve the high-performance demand of network IDS. Likewise, the authors in \cite{8418451} have studied the effectiveness of a stacked sparse AE model for extracting useful features of intrusion behavior. The study results indicated that the model can extract more discriminative features and accelerate the detection process. Relatedly, the authors in  \cite{kim2019designing} proposed new interesting online deep learning systems that apply an AE as function approximation in the Q-network of RL to achieve a higher detection accuracy rate for network intrusion detection.\\
From the above literature review, it is evident that despite the significant performance gain achieved with the application of unsupervised approaches in IDS design, there is still room for improvement. one of the causes of weakness include a focus on the last snapshot, while there are temporal and spatial dependencies in the traffic data. On these grounds, the existing approaches are prone to overfit and show poor generalization performance toward unseen cyberattacks. Thus, research on unsupervised IDS is still in its infancy in terms of development. Hence, the proposed research is expected to make a valuable contribution to the existing knowledge pool.

\section{Preliminaries}
The approach presented in our study makes use of different typologies of neural networks arranged to provide a powerful network traffic classifier suitable for most of the tasks characterizing modern ISP activities. In this section, the basic theoretical backgrounds of the employed networks are presented.
\subsection{Autoencoder}
An Autoencoder (AE) is an unsupervised feed-forward neural network used for the reconstruction of its input. AE attempts to find an optimal subspace where the normal data and anomalous data appear very different.
Let us assume that the normal training set is $\{x_1, x_2, x_3,...,x_n\}$, each of which is a $d$ dimensional vector $(x_i \in R^d)$. In the training phase, we construct a model to project these training data into the lower-dimensional subspace and reproduce the data to get the output $\{x_{1}^{\prime},x_{2}^{\prime}, ...,x_{n}^{\prime}\}$. Therefore, we optimize the model to minimize reconstruction error so as to get the optimal subspace. The reconstruction error is defined as:
\begin{equation} \label{eq:err}\varepsilon(x_{i},\ x_{i}^{\prime})=\sum_{j=1}^{d}(x_{i}-x_{i}^{\prime})^{2} \tag{1} \end{equation}

As the normal data in the test dataset meet the normal profile which is built in the training phase, the corresponding error is smaller, whereas the anomalous data will have a relatively higher reconstruction error. As a result, by thresholding the reconstruction error, we can easily classify the anomalous data:

$$ c(x_{i})=\left\{
\begin{aligned}
normal  \varepsilon_{i} < \theta) \\
anomalous \varepsilon_{i} > \theta
\end{aligned}
\right.
$$

The architecture of the Autoencoder consists of the encoder and the decoder. The encoder and decoder are composed of an input layer, an output layer, and one or more hidden layers. It has a symmetrical pattern – the output layer of the decoder is equal to the input layer of the encoder.
Mathematically, encoder  with input vectors $(x_i \in R^d)$ and output layer of size $m$ (hidden layer) can be defined as :
\begin{equation*} h_{i}=f_{\theta}(x)=s(\sum_{j=1}^{n}W_{ij}^{input}x_{j}+b_{i}^{input}) \tag{3} \end{equation*}

where $x$ is the input vector, $\theta$ is the parameters $\{W^{input}, b^{input}\}$, $W$  represents the encoder weight matrix with size $m \times d$ and $b$ is a bias vector of dimensionality $m$. Therefore, the input vector is encoded to a lower-dimensional vector. The resulting hidden representation $h_i$ is then decoded back to the original input space $R^d$ using a decoder. The mapping function is as follow:
\begin{equation*} x_{i}^{\prime}=g_{\theta^{\prime}}(h)=s(\sum_{j=1}^{n}W_{ij}^{hidden}h_{j}+b_{i}^{hidden}) \tag{4} \end{equation*}

The parameter set of the decoder is $\theta^\prime = \{W^{hidden}, b^{hidden}\}$. We optimize the autoencoder to minimize the average construction error with respect to $\theta$ and $\theta^\prime$:
\begin{align*} \theta^{\ast}, \theta^{\prime\ast}&=\substack{\arg \min\\ \theta,\theta^{\prime}}\frac{1}{n}\sum_{i=1}^{n}\varepsilon(x_{i}, x_{i}^{\prime})\\ &=\substack{\arg\min\\ \theta,\theta^{\prime}}\frac{1}{n}\sum_{i=1}^{n}\varepsilon(x_{i},\ g_{\theta^{\prime}}(f_{\theta}(x_{i}))) \tag{5} \end{align*}

where $\epsilon$ is the reconstruction error in Equation (1). After we finish training this Autoencoder, we can feed the test data into it to compute the reconstruction errors for each set of data. The anomalous data can be determined by utilizing equation (3). To be noted here, the activation functions $f$ and $g$ should be non-linear functions so as to reveal the non-linear correlation between the input features.\\

\subsection{Isolation Forest}
Isolation Forest \cite{sadaf2020intrusion} is an unsupervised machine learning method that can find anomalies by randomly partitioning the data points. Isolation Forest assumes that the instances which fall away from the data center are anomalies. It forms like binary trees and ensembles iTrees by sampling randomly for a given dataset. The isolation tree’s key role is to make use of unusual samples, also called anomalies in detecting the unknown attacks which are strange from the normal attacks. Random selection of a subset from the training set is done to build iTrees and it was found that the realistic amount is $256$ after subsampling and this is the first step in creating iForest. It does not make use of distance measures, hence reduction found in the cost required for computing. Secondly, iForest utilizes no distance or density measures to detect an anomaly, this second step, therefore, eliminates computational cost compared with distance measures involved in clustering and takes time complexity in a linear fashion. Lastly, iForest requires a low amount of memory and uses the idea of ensemble and does not bother if some iTrees does not yield efficient results as the ensemble algorithms convert the weak trees into efficient ones. Due to all these benefits, using iForest is strongly recommended to detect anomalies on huge datasets involving complex features. It calculates anomaly score S as
\begin{equation} \tag{6}
S=2^{-\frac {E(h\left ({x }\right))}{C(n)}}
\end{equation}
where $h(x)$ is the number of edges in a tree for a certain point $x$ and $C(n)$ is normalization constant for a dataset of size $n$ . The binary class gets separated based on a threshold value on anomaly score in supervised classification and without threshold value in unsupervised classification.

\begin{figure*}[ht]
    \centering
    \includegraphics[scale=0.5]{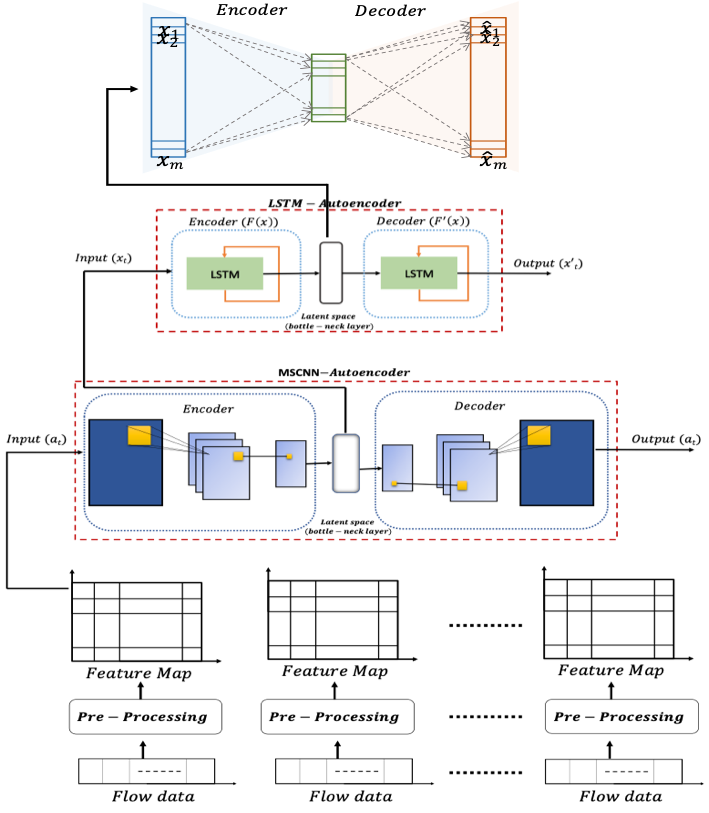}
    \caption{Spatio-temporal features integration in the architecture of the MSCNN-LSTM-AE model}
    \label{fig:models}
\end{figure*}
\begin{figure*}[ht]
    \centering
    \includegraphics[scale=0.5]{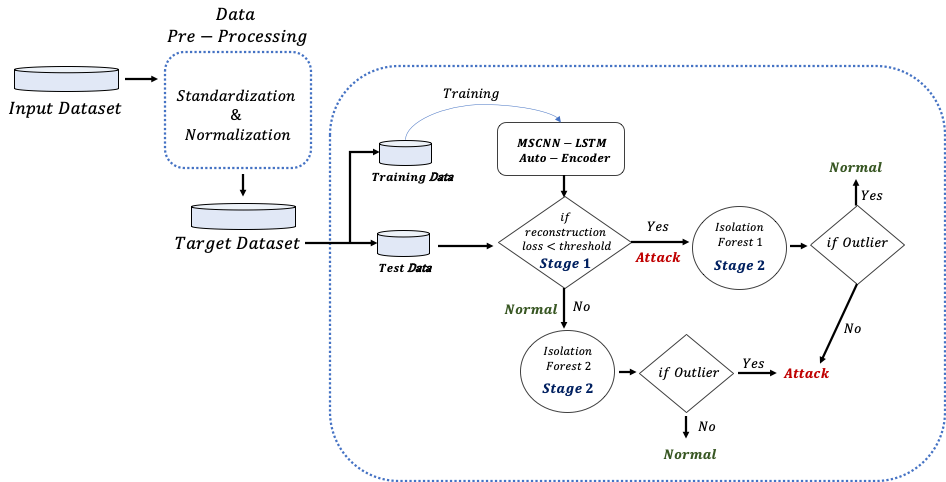}
    \caption{A block diagram of MSCNN-LSTM-AE model learning process}
    \label{fig:flowmodels}
\end{figure*}
\section{The Proposed methodology}
Our proposed approach consists of data preprocessing and spatio-temporal feature integration using multi-scale CNN-AE and LSTM-AE as shown in Figure \ref{fig:models}. Our model computes anomaly scores based on the reconstruction error for traffic data to identify malicious traffic.
Our proposed method involves two stages of anomaly detection. The output of the first stage acts as the input to the second stage. As depicted in the flowchart shown in Figure \ref{fig:flowmodels}, the test dataset is supplied to the autoencoder (MSCNN-LSTM) in stage 1. Unified MSCNN-LSTM-AE identifies the attack based on threshold and segregates the attack and normal network traffic data into two sets. However, the resultant sets contain data points that ideally don’t belong to them. Isolation forest in stage 2, attempt to identify these misfit (outlier) data points, which improves the overall accuracy.
\subsection{Data Pre-processing}
There are some features that are symbolic and continuous. These features need to be converted into a single numeric type for feature extraction. Secondly, features are not uniformly distributed thus need to be scaled for a better result with machine learning models.
Data standardization in pre-processing deals with the numeralization of categorical features. The most common method is encoding symbolic values with numeric values. For example, if feature contain three unique symbolic values like $protocol$ $type$ contain $tcp$, $udp$ and $icmp$ then these attributes can be map with 1, 2 and 3 respectively.
Generally, features in traffic data flow are highly variable and not uniformly distributed. To achieve better results with the machine learning model, the attribute values are usually scaled to a uniform distribution in the interval $[0-1]$. For this purpose, the min-max data normalization method is used, as shown in Equation $7$.


\begin{equation} \tag{7}
	x_{i}^{'} = \frac{x_{i} - min (x_i)}{max(x_i) - min(x_i)}
\end{equation}

Where $max(x_i)$ and $min(x_i)$ represent maximum and minimum value of feature vector $x_i$; whereas $x_{i}^{'}$ is a normalized feature value between [0-1].\\
Usually, the first order features captured from network traffic are arranged as a
dimensional (1D) vector. Let d be such a dimension, the d‐dimensional input needs to be arranged to a 2D matrix for being fed into 2D‐CNN. Accordingly, let $x \in \mathbb{R}^d$ be the  pre-processed input and let $ d^{'} = (N_x \times N_y)$ be a derived 2D‐matrix (feature maps) as shown in Figure \ref{fig:models}.
\subsection{Multi-scale CNN-AE based spatial feature extraction}

The convolutional neural network (CNN) architecture performs well in the image processing field. However, during the image processing, CNN focuses on some local features of the image such as edge information. The identification of network traffic can not only rely on some discrete local features but also need to combine multiple local features to perform the classification. Therefore, the CNN is adjusted and transformed into Multi-scale CNN to accomplish this task. When a human visual perception system maps an image in the brain, it will first form a complete set of images from far to near, and from fuzzy to clear. Therefore, the MSCNN simulates different projections of objects at different distances on the retina during human eye recognition. Similarly, network traffic is a high-dimensional dataset that cannot be identified by only a few discrete features. \\
Convolutional Autoencoder (CAE) \cite{yu2017network} is a special kind of autoencoder that does not apply the fully connected neural layer. The CAE model consists of convolutional and deconvolution layers from CNN architecture. It utilizes a convolutional layer in the encoder part and a deconvolutional layer in the decoder part. Apparently, the convolutional layer is able to decrease the feature number while the deconvolution layer can increase the feature number. As a result, in CAE, the convolutional layer takes the role of the encoder to perform the dimensionality reduction, while the deconvolution layer is applied here to reconstruct the data. CAE takes advantage of the Convolutional and Deconvolution layers. As these layers utilize the kernel filters, a parameter sharing scheme is applied here to control the number of parameters. Therefore, compared with conventional Autoencoder, CAE has a smaller number of parameters so the training time of CAE is much smaller.\\
In the MSCNN-AE, we used multiple convolution kernels of different sizes to extract feature maps and combine them to obtain multiple sets of local features to achieve accurate identification. The MSCNN structure is based on three original multi-scale convolutional as shown in Figure \ref{fig:models}. The multi-scale convolution layer extracts features of the dataset using $1*1$, $2*2$, and $3*3$ convolution kernels. The aim of this configuration is to mine spatial features expressing relevant relations among the basic features. With reference to Figure \ref{fig:models}, the encoder and decoder of AE include input layer, convolutional layers, pooling layer, fully-connected NN (latent space), and output layer equal to input layer size.\\
Let $X \in \mathbb{R}^{N_x \times N_y}$ and $K^f \in \mathbb{R}^{a \times b}$ be the CNN-input, and the filter, respectively. The convolution operation between the CNN-input
$X$ and $N_f$ filters is defined by:
\begin{equation} \tag{8}
	Y_{i,j} = \sum_{f=1}^{N_f}\sum_{p=1}^a\sum_{q=1}^b K_{p,q}^{f}X_{i+p-1,j+q-1}
\end{equation}
with $Y_{i,j}$ the components of the filtered input. The size of Y is defined through its row $(Y_x)$ and column $(Y_y)$ dimension, by:
\begin{equation} \tag{9}
	Y_{x} = \frac{N_x - a +2P}{S_x}+1
\end{equation}
\begin{equation}\tag{10}
	Y_{y} = \frac{N_y - a +2P}{S_y}+1
\end{equation}
where $S_x$ and $S_y$ are the Strides on the row and column, respectively,
which control the shifting of the filter on the input. In addition, $P$ is
the Padding, which controls the number of zeros around the border of $X$. Padding is used to change the size of CNN-output without compromising the convolution result.
By assuming $d = N_x \times N_y$, it is possible to map any $x_k$ to a point $X_{i,j}$ in a two-dimensional array (that looks like a $N_x \times N_y$ matrix). Thus, with abuse of notation, we can assert that: $x_k \equiv X_{i,j}$ with k = 1...d, i=1...$N_x$, and j= 1...$N_y$.
\begin{equation}\tag{11}
h_l = \sigma\left(\sum_{k=1}^{d^{'}}w_{lk}^{'}x_{\phi{k}}+b_l\right)
\end{equation}
with $d^{'}= Y_{x} \times Y_{y}$, while
\begin{equation}\tag{12}
w_{lk}^{'} = \sum_{k=1}^{N_f}\sum_{p=1}^{a}\sum_{q=1}^{b}K_{p,q}^{f}w_{lk}
\end{equation}
and $x_{\phi{k}} \equiv X_{{i+p-1},{j+q-1}}$ with $\phi(k) = ({i+p-1},{j+q-1})$. The $w_{ik}^{'}(\forall l,k)$
represent the new extracted features and act as input to the LSTM Autoencoder. This latent space features express a more complex knowledge because they are a linear combination of the original.\\
\begin{figure}[ht]
    \centering
    \includegraphics[scale=0.4]{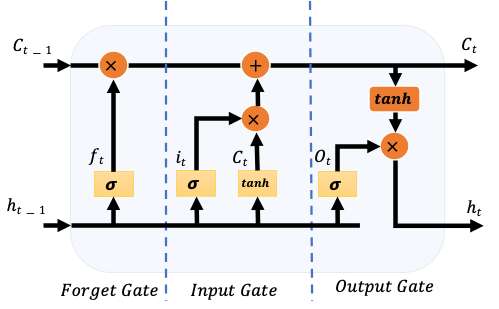}
    \caption{Schematic representation of a LSTM cell}
    \label{fig:lstm1}
\end{figure}

\subsection{LSTM-AE based feature extraction}
LSTM based Autoencoder models are used to form a sequence to sequence architecture. LSTM-AE consist of encoder and decoder function. An encoder function task is learning the prominent characteristics and creating an encoded version of the input sample. The decoder aims to reconstruct the input using internal representation.\\
In our case, LSTM-AE architecture's encoder function maps a sequence of latent space features extracted from network traffic (high-level features) by MSCNN-AE into a fixed-length vector of new features (latent space) which in turn is converted to the same input sequence using decoder function. This configuration is able to mine short and long-distance dependencies within the sequence of basic-features  
LSTM can remember long-term dependency using "LSTM Cell" $(c_t)$, which allows the cell to remember or forget past information (shown in Figure \ref{fig:lstm1}) . This state is updated through four internal activation layers called gates, implemented by a sigmoid neural network layer and a point-wise operation. Each gate is devoted to a specific goal, as described in Equations 13,14, 15, 16, 17, and 18 below:
\begin{equation}\tag{13}
f_t = \sigma(W_f.[h_{t-1},x_t]+b_f)
\end{equation}
\begin{equation}\tag{14}
	i_t =\sigma(W_i.[h_{t-1},x_t]+b_i)
\end{equation}
\begin{equation}\tag{15}
	\Tilde{C}_t=tanh(W_c.[h_{t-1},x_t]+b_c)
\end{equation}
\begin{equation}\tag{16}
	O_t = \sigma (W_o[h_{t-1}, x_t] + b_o)
\end{equation}
\begin{equation}\tag{17}
	{C}_t = f_t * C_{t-1} + i_t * \Tilde{C}_t
\end{equation}
\begin{equation}\tag{18}
	h_t =O_t *tanh(C_t)
\end{equation}
where $W_f,W_i, W_c , W_o$ are linear transformation and $C_t,h_t$ are cell memory and output at time $t$.\\
LSTM-AE receives a traffic sequence of length W where $n$ is normal training example index such that $X(n) = [x^{(1)},x^{(2)},...,x^{(W)}]$. The LSTM-encoder produces a synthesized output-vector $(y^W)$ of a pre-determined $r \times 1$ dimension based on  equations from 13 to 18, which can be expressed by:
\begin{equation}\tag{19}
	y^W = \psi(x^{(1)},x^{(2)},x^{(3)},...,x^{(W)})
\end{equation}
where $\psi$ represent non-linear encoder funtion of LSTM architecture. This $y^W$ is new latent-space features express a compact representation of the temporal behavior of the basic-features. This $y^W$ is used by decoder to reconstruct input sample as depecited by Equation 19:
\begin{equation}\tag{20}
	\hat {\boldsymbol {X(n)}} = \Phi(y^{(1)},y^{(2)},y^{(3)},...,y^{(r)})
\end{equation}
where  $\Phi$ is the decoder function of LSTM autoencoder (LSTM-AE). The objective of the decoder is to reconstruct input sequence with minimum loss whereas loss is calculated in terms of mean square error as shown in Equation ~\eqref{eq:err}

\subsection{Classification}
We train our Autoencoder model on latent space features based on normal data extracted from LSTM AE  as shown in Figure \ref{fig:flowmodels}. Since the AE is only trained on "normal" data, the reconstruction loss for the attack data is much higher than the normal data. 
For example, if the reconstruction loss value of a data point is higher than the threshold value, then data point is classified as "attack", otherwise, it will be classified as "normal".
\begin{equation}\tag{21}
	Y_{i}^{'} = ({(Y_{p}^{'}),(Y_{q}^{'})})
\end{equation}
where $Y_{p}^{'}$ represents "normal" data packets having reconstruction loss error less than the threshold and $Y_{q}^{'}$  contains data points having reconstruction error greater than the threshold and are considered as "attacks". Since the result of AE is not hundred percent accurate, both $Y_{p}^{'}$ and $Y_{q}^{'}$ contain attack and normal data respectively.\\
To achieve more accuracy i.e. detecting more intrusions,
these 2 sets are then supplied as inputs to two Isolation Forest
modules (Figure \ref{fig:flowmodels}). The first module (Isolation Forest 1) gets the "attack" output of the AE and search for the anomalies, in our case-normal data points. Similarly, the second module
(Isolation forest 2) takes the "normal" output of the AE and search for anomalies, in this case - attack data points. The attack data in the "normal" set and normal data in the "attack" set are nothing but outliers or anomalies. Isolation Forest 2 takes the "normal" $Y_{p}^{'}$ and searches for attack data. Since AE has already identified most of the normal and attack packets in stage 1, the set $Y_{p}^{'}$ contains a fewer number of attack packets. The set $Y_{q}^{'}$ containing "attack" data is fed to Isolation Forest 1. $Y_{q}^{'}$ contains some actual normal data too. Isolation Forest 1 searches for these "outliers" in $Y_{q}^{'}$.

\begin{equation}\tag{22}
	Y_{p}, O_{q} \leftarrow Isolation Forest 1((Y_{p}^{'}))
\end{equation}
\begin{equation}\tag{23}
	Y_{q}, O_{p} \leftarrow Isolation Forest 2((Y_{q}^{'}))
\end{equation}
At the end, $(Y_{p}, O_{p})$ and $(Y_{p},O_{q}) $ are final set of normal and malicious network traffic packets.

\begin{figure*}

\begin{tabular}{ccc}
  \includegraphics[width=55mm]{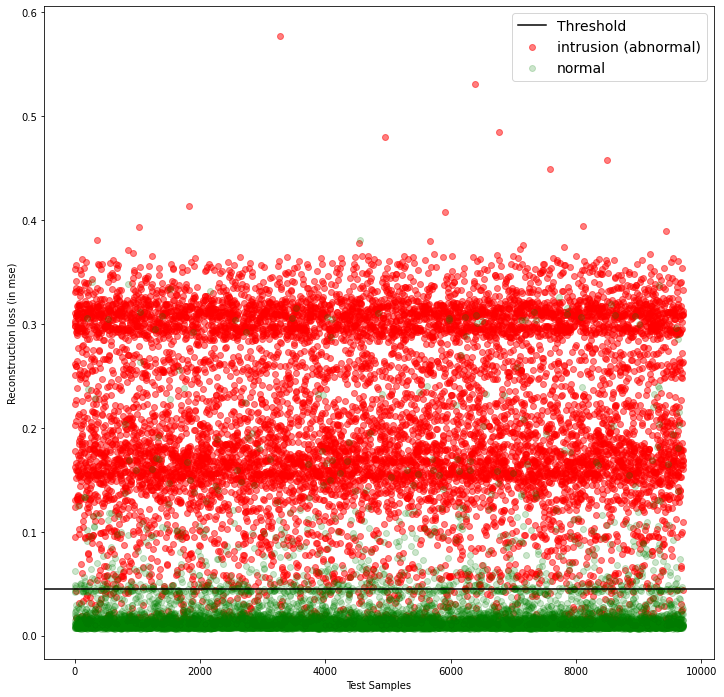} &   \includegraphics[width=55mm]{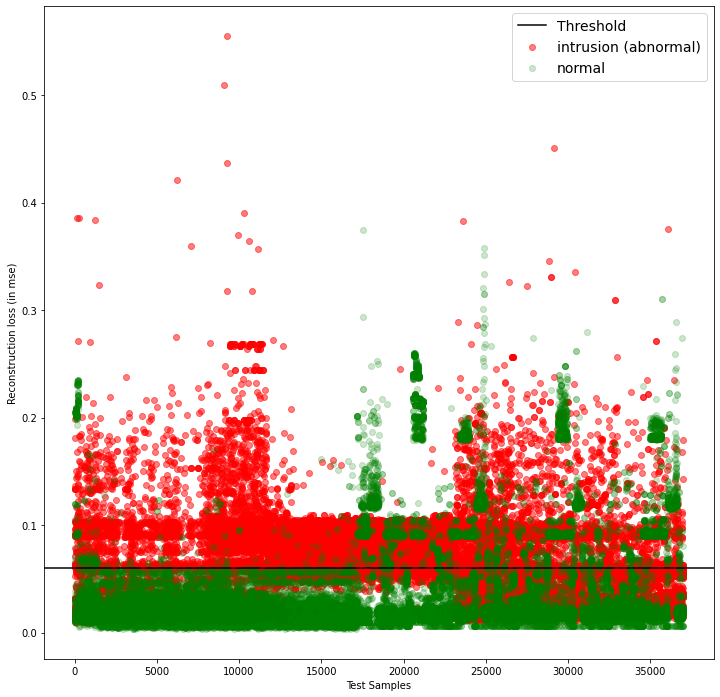}  & \includegraphics[width=55mm]{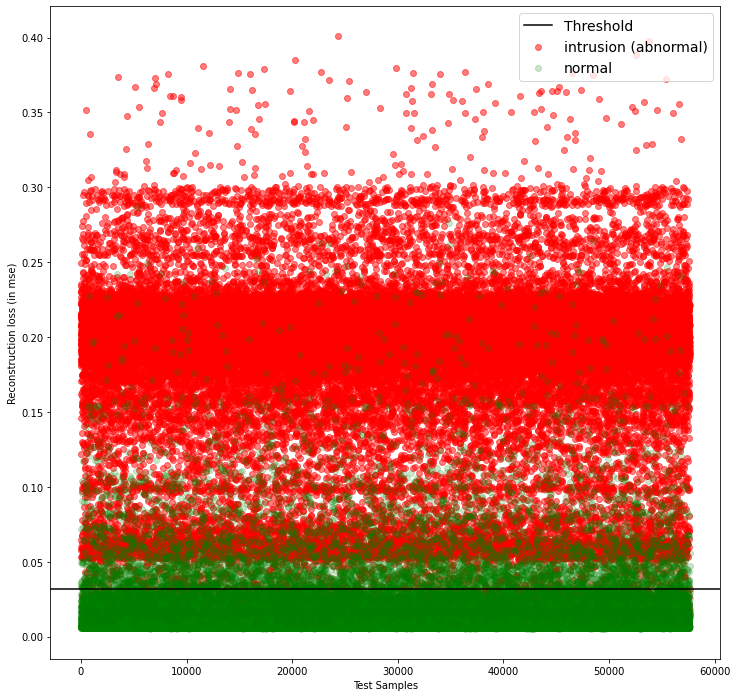}\\
(a) NSL-KDD  &  (b) UNSW-NB15  & (c) CICDDoS2019\\[3pt]
\end{tabular}
\caption{Anomaly score based on reconstrusion error in NSL-KDD, UNSW-NB15 and CICDDoS2019 test datasets respectively based on MSCNN-LSTM autoencoder.}
\label{fig:mse}
\end{figure*}

 \section{Experiment setup and evaluation metrics}
This section describes the dataset, experiment setup, performance metrics, and the dataset used to evaluate the proposed approach.
\subsection{Dataset Description}
To evaluate our proposed approach, we considered well-known public datasets namely, NSL-KDD \cite{nslkdd}, UNSW-NB15 \cite{unswnb} and CICDDoS2019 respectively.
This benchmark dataset is freely available from the Canadian Institute of Cybersecurity. NSL-KDD dataset was obtained by removing redundant records from the KDDCUP99 dataset so that machine learning-based models can produce unbiased results. In addition to normal traffic, this dataset consists of traffic from four attacks, namely DoS, U2R, R2L, and PROBE traffic.


\begin{table}[ht]
\centering
\caption{NSL-KDD dataset}
\label{table:nslkdd2}
\scalebox{0.90}{
\begin{tabular}{@{}lcc@{}}
\toprule
$Class$  & NSL-KDD$_{Train+}$ & NSL-KDD$_{Test+}$ \\ \midrule
$Normal$  & 67343     & 9711     \\
$Attack$  & 57738     & 12833     \\
$Total$  & 125081     & 22544     \\ \bottomrule
\end{tabular}}
\end{table}

Table ~\ref{table:nslkdd2} shows the count of samples from the training and testing set from the NSL-KDD dataset. As shown in Table ~\ref{table:nslkdd2}, it is highly imbalanced with fewer instances from U2R and R2L attack classes. Furthermore, the test dataset contains unknown attacks samples that do not appear in the training dataset.

Each traffic record in the NSL-KDD dataset is a vector of 41 continuous and nominal values. These 41 values can be further subdivided into four categories. The first category is the intrinsic type, which essentially refers to the inherent characteristics of an individual connection. The second category contains indicators that relate to the content of the network connection. The third category receives a set of values based on the study of the content of the connections in the time segment of 2 seconds. Finally, the fourth category is based on the destination host. \\
UNSW-NB15 dataset is a benchmark dataset that contains nine families of intrusion attacks, namely, Shellcode, Fuzzers, Generic, DoS, Backdoors, Analysis, Generic, Worms, and Reconnaissance \cite{unswnb}. This dataset is freely provided by the Cyber Range Lab of the Australian Centre for Cyber Security (ACCS). We used an already configured training and testing data set from ACCS as shown in Table \ref{table:UNSW}. The number of samples in the training dataset is 175,341 and the test set has 82,332 from normal and nine types of attacks. The dataset has a total of 42 features and these features can be subdivided into categories. Similar to the NSL-KDD dataset, the first part is the content features; the second category has some features which refer to the basic and general operation of the internet; the third part is connection features and lastly, the fourth category is time-based features.

\begin{table}[ht]
\centering
\caption{UNSW-NB15 dataset}
\label{table:UNSW}
\scalebox{0.90}{
\begin{tabular}{@{}lcc@{}}
\toprule
$Class$  & UNSW-NB15$_{Train}$ & UNSW-NB15$_{Test}$ \\ \midrule
$Normal$  & 56000     & 37000     \\
$Attacks$    & 119341    & 45332     \\
$Total$    & 175341    & 82332    \\ \bottomrule
\end{tabular}}
\end{table}
Another dataset, we used is the CICDDoS2019 dataset that has been widely used for DDoS attack detection and classification. The dataset contains a large amount of up-to-date realistic DDoS attack samples as well as benign samples. The total number of records contained in CICDDoS2019 is depicted in Table \ref{table:no_dataset}. We have used all the benign samples and $1\%$ malicious samples from training and test data sets.
\begin{table} [h]
\setlength{\tabcolsep}{3.5mm}
\caption{The number of records in CICDDoS2019 }
 	\label{table:no_dataset}
  \begin{tabular}{cccc}
  \midrule
   {\textbf{dataset}} &
   {\textbf{total}} &
   {\textbf{benign}} & {\textbf{malicious}} \\
 \hline
Training day & 50,063,112	& 56,863	& 50,006,249 \\
Testing day & 20,364,525    & 56,965    & 20,307,560 \\
 \midrule
 \end{tabular}
\end{table}
Each record of the dataset contains $87$ statistical features (e.g., timestamp, source, and destination IP addresses, source, and destination port numbers, the protocol used for the attack, and a label for a type of DDoS attack). The training dataset contains a total of 12 different types of DDoS attacks (i.e., NTP, DNS, LDAP, MSSQL, NetBIOS, SNMP, SSDP, UDP, UDP-Lag, WebDDoS, SYN, and TFTP) while only 7 DDoS attacks are included in the testing dataset (i.e., PortScan, NetBIOS, LDAP, MSSQL, UDP, UDP-Lag and SYN).

\subsection{Experimental Setup}
This study was carried out using a 2.3 GHz 8-core Intel i9 processor with 16 GB memory on MacOS Big Sur 11.4 operating system. The proposed approach is developed using Python programming language with several statistical and visualization packages such as Sckit-learn, Numpy, Pandas, Tensorflow, and Matplotlib. Table~\ref{table:Mat} summarizes our system configuration.
\begin{table}[ht]
	\centering
	\footnotesize
	\caption{Implementation environment specification}
	\label{table:Mat}
	\begin{tabular}{p{2.6cm} | p{3.8cm}}
		\hline
		\textbf{Unit}   & \textbf{Description}\\ \hline
		Processor   & 2.3 GHz 8-core Inter Core i9 \\ \hline
		RAM  &  16 GB      \\ \hline
		Operating System  &  MacOS Big Sur 11.4  \\ \hline	
		Packages   &  Tensorflow, Sckit-Learn, Numpy, Pandas, Pyriemannian and Matplotlib    \\ \hline
	\end{tabular}
\end{table}
\begin{figure*}
\begin{tabular}{ccc}
  \includegraphics[width=55mm]{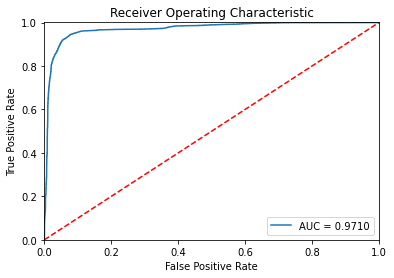} &   \includegraphics[width=55mm]{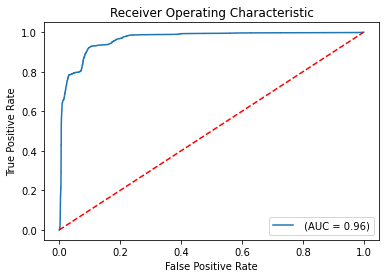}  & \includegraphics[width=55mm]{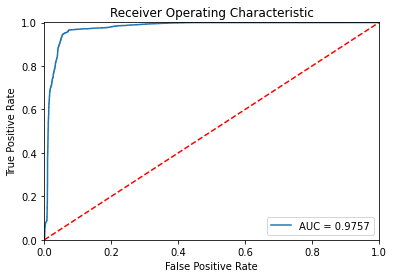}\\
(a) NSL-KDD  &  (b) UNSW-NB15  & (c) CICDDoS2019\\[3pt]
\end{tabular}
\caption{Reciver operting curve showing area under curve for our proposed model.}
\label{fig:roc}
\end{figure*}
\subsection{Evaluation Metrics}
The proposed method is compared and evaluated using Accuracy, Precision, Recall, F1-score, and Area under the receiver operating characteristics (ROC) curve. In this work, we have used the macro and micro average of Recall, Precision, and F1-score for multi-class classification. All the above metrics can be obtained using the confusion matrix (CM). Table ~\ref{tab:conmat} illustrates CM for binary classes but can be extended to multiple classes.\\

\begin{table}[ht]
\centering
\caption{Illustration of confusion matrix}
\label{tab:conmat}
\scalebox{0.85}{
\begin{tabular}{@{}cccc@{}}
\toprule
\multicolumn{1}{l}{}    &         & \multicolumn{2}{c}{Predicted}        \\\cmidrule(l){2-4}  
\multicolumn{1}{l}{}    &         & Class$_{pos}$ & Class$_{neg}$ \\ \midrule
\multirow{2}{*}{Actual} & Class$_{pos}$ & True Positive $(TP)$     & False Positive $(FP)$      \\
                        & Class$_{neg}$ & False Negative $(FN)$     & True Negative $(TN)$      \\
\bottomrule
\end{tabular}}
\end{table}
\begin{figure*}
\begin{tabular}{ccc}
  \includegraphics[width=55mm]{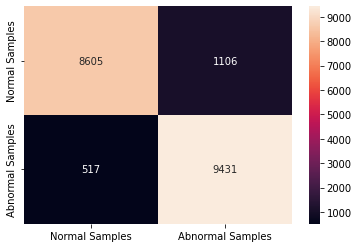} &   \includegraphics[width=55mm]{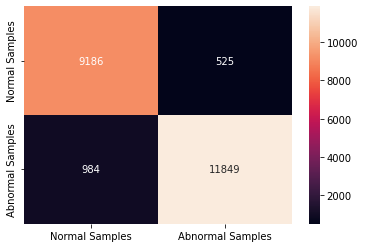}  & \includegraphics[width=55mm]{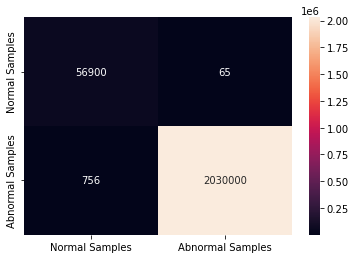}\\
(a) NSL-KDD  &  (b) UNSW-NB15  & (c) CICDDoS2019\\[3pt]
\end{tabular}
\caption{Confusion matrix for NSL-KDD, UNSW-NB15 and CICDDoS2019 test datasets respectively.}
\label{fig:cm}
\end{figure*}
In Table~\ref{tab:conmat}, True positive (TP) means amount of class$_{pos}$ data predicted actual belong to class$_{pos}$, True negative (TN) is amount of class$_{neg}$ data predicted is actually class$_{neg}$, False positive (FP) indicates data predicted class$_{pos}$ is actual belong to class$_{neg}$ and False negative (TN) is data predicted as class$_{neg}$ but actually belong to class$_{pos}$. Based on the aforementioned terms, the evaluation metrics are calculated as follows. \\
Accuracy (ACC) measures the total number of data samples are correctly classified as shown in equation 24. For balanced test dataset, higher accuracy is indicate model is well learned, but for unbalanced test dataset scenarios relying on accuracy can give wrong illusion about model's performance.
\begin{equation} \tag{24}
	ACC = \frac{TP+TN}{TP + TN + FP + FN}
\end{equation}
Recall (also known as true positive rate) estimates the ratio of the correctly predicted samples of the class to the overall number of instances of the same class. It can be computed using equation 25 . Higher Recall $\in [0,1]$ value indicate good performance of machine learning model.
\begin{equation} \tag{25}
	Recall = \frac{TP}{TP + FN}
\end{equation}
Precision measures the quality of the correct predictions. mathematically, it is the ratio of correctly predicted samples to the number of all the predicted samples for that particular class as shown in Equation 26. Precision is usually paired with Recall to evaluate the performance of the model. Sometimes pair can appear contradictory thus comprehensive measure F1-score is considered for unbalanced test data-sets.
\begin{equation}\tag{26}
	Precision = \frac{TP}{TP + FP}
\end{equation}
F1-Score computes the trade-off between precision and recall. Mathematically, it is the harmonic mean of precision and recall as shown in equation 27.
\begin{equation}\tag{28}
	F = 2\times\left(\frac{Precision\times Recall}{Precision + Recall}\right)
\end{equation}
The area under the curve (AUC) computes the area under the receiver operating characteristics (ROC) curve which is plotted based on the trade-off between the true positive rate on the y-axis and the false positive rate on the x-axis across different thresholds. Mathematically, AUC is computed as shown in Equation 29.
\begin{equation}\tag{29}
AUC_{ROC}=\int_{0}^{1} \frac{TP}{TP+FN}d\frac{FP}{TN+FP}
\end{equation}
In the case of unbalanced test data classification, the performance of models is usually evaluated using macro and micro-averaging of recall, precision, and F1-score. Macro-averaging in simple terms is the arithmetic mean of the individual precision, recall, and F1-scores while micro-averaging sums up the individual TP's, FP's, and FN's.


\section{Results and Discussion}
We have made a number of different observations to understand the performance implications both during the training and testing phases.
Figure \ref{fig:mse} shows anomaly score in terms of reconstruction error for each data point in the NSL-KDD  test dataset based on our model. Here, we can clearly see that some data points in normal and abnormal traffic are wrongly classified based on a threshold. In our case, any point whose reconstruction error is mean plus two standard deviations from normal training samples reconstruction loss is classified as an abnormal data point. To rectify the issue of misclassification based on the threshold we have employed second stage classification based on isolation forest that removes outlier data points from the abnormal and normal set made after Autoencoder's threshold-based classification. Table \ref{table:Comparison} shows the performance of our proposed (MSCNN-LSTM-AE with isolation forest)  approach on the NSL-KDD dataset able to achieve  93.76 percent accuracy and 92.26 percent recall respectively. In addition to this, we compared our method's performance with other state-of-the-art methods using the four metrics namely accuracy, precision, recall, and F1-score.  As the results in the table show our approach obtained high recall and F-score respectively compared to all the state-of-the-art methods in the literature. High recall and F-score indicate better performance of the model.

Similarly, We also bench-marked our model on UNSW-NB15 and CICDDoS2019 datasets. Figure \ref{fig:cm} shows confusion matrix visualization of NSL-KDD, UNSW-NB15, and CICDDoS2019 dataset obtained through our model. As we can see, still some data points are wrongly classified that we tried to correct through second stage classification by isolation forest as shown in Table \ref{table:Comparison}. Our model obtained high recall and F-score respectively compared to other state-of-the-art methods. In this work, we do not check the effects of different hyper-parameter, hidden layers, and loss functions on the performance of our model but  If we fine-tune our model's parameters with some optimization technique, we can further increase the performance of our intrusion detection models which is mostly the case with other state-of-the-art methods in the literature \cite{article4}.



\begin{table*}[h]
	\centering
	\footnotesize
	\footnotesize
	\footnotesize
	\caption{Comparison to other similar methods}
	\label{table:Comparison}
	\begin{tabular}{c|c|c|c|c|c|c}
		\midrule
		Paper      & Dataset & Techniques       & Acc & Precision & Recall  & F1-score \\ \hline
			Sharafaldin et al.~\cite{sharafaldin2019developing} & CICDDoS2019 &Random Forrest              & -   & 77    & 56 & 62
		\\ \hline
		Rajagopal et al.~\cite{rajagopal2021towards}& CICDDoS2019   & Extended Decision Tree & 97       & 99.0        & 97.0      & 97.8
			\\ \hline
		Gohil et al.~\cite{gohil2020evaluation} & CICDDoS2019 & Extended Naive Bayes             & 96.25   & 96    & 96 & 96
		\\ \hline
	
		Shieh et al.~\cite{shieh2021detection}& CICDDoS2019   & Bi-LSTM & 98.18       & 97.93        & 99.84      & -
			\\ \hline
		De Assis et al.~\cite{de2020near}& CICDDoS2019  & CNN              & 95.4   & 93.3    & 92.4 & 92.8 \\ \hline
		De Assis et al.~\cite{de2020near}& CICDDoS2019  & MLP              & 92.5   & 84.4    & 94.2 & 89.0 \\ \hline
Javaid et al. \cite{deep2}& CICDDoS2019  & AE + Regression & 88.39 & 85.44 & 95.95& 90.4\\ \hline
Sadaf et al. \cite{9189883}& CICDDoS2019  & AE + Isolation Forest & 88.98 & 87.92 & 93.48 & 90.61 \\\hline
Can et al.\cite{can2021detection} & CICDDoS2019 & FS + MLP & - & 91.16 & 79.41 & 79.39 \\ \hline
Wei et al.\cite{9591559}& CICDDoS2019  & AE + MLP & 98.38 & 97.91 & 98.48 & 98.18 \\ \hline
		\textbf{Our method} & CICDDoS2019 & MSCNN-LSTM-AE         & 99.56    & 98.91     & 98.81   & 98.46   \\ \midrule
		B. Ingre\cite{7058223}& NSL-KDD & ANN  & 81.2 & \textbf{96.59} & 69.35 & 80.73\\\hline
M. Al-Qatf\cite{8463474}& NSL-KDD  & Sparse-AE + SVM  & 84.96 & \textbf{96.23} & 76.57 & 85.28 \\\hline
I. Sharafaldin\cite{IERACITANO202051}& NSL-KDD &AE   & 84.21 & 87 & 80.37 & 81.98 \\\hline
I. Sharafaldin\cite{IERACITANO202051}& NSL-KDD &LSTM  & 82.04 & 85.13 & 77.70& 79.24\\ \hline
I. Sharafaldin\cite{IERACITANO202051}& NSL-KDD &MLP  & 81.65 & 85.03 & 77.13 & 78.67 \\\hline
Sadaf et al. \cite{9189883}& NSL-KDD & AE & 88.98 &87.92& 93.48 & 90.61 \\\hline
Javed et al.\cite{7966342}& NSL-KDD  & AE & 88.39 & 85.44& 95.95 & 90.04\\\hline
Wen et al.\cite{9552882}& NSL-KDD & AE  & 90.61 & 86.83 & 98.43 & 92.26\\\hline
\textbf{Our method}& NSL-KDD & MSCNN-LSTM-AE         & \textbf{93.30} & 95.75 & \textbf{92.33} & \textbf{94.01}\\ \hline
H. Zhang\cite{article4}& UNSW-NB15 &AdaBoost  & 86.41 & 72.83 & 95.96 & 82.81\\\hline
H. Zhang\cite{article4}& UNSW-NB15  &Extra Trees & 86.73 & 72.60 & 97.14 & 83.10 \\ \hline
Kasongo and Sun \cite{kasongo2020performance}& UNSW-NB15  & ANN & 86.71 & 81.54 & 98.06 & 89.04 \\\hline
Hammad et al. \cite{9312002}& UNSW-NB15 	& Naive Bayes & 76.04 & 76 & 83.4 & 76.8\\	\hline
Dickson and Thomas \cite{10.1007/978-981-16-0419-5_16}& UNSW-NB15  & Logistic regression &  84 & - & - &- \\ \hline
Amaizu et al. \cite{9289329}& UNSW-NB15  & DNN & 88 & 90 &  88 & 87 \\ \hline
\textbf{Our method}& UNSW-NB15  & MSCNN-LSTM-AE         & \textbf{89} & \textbf{88} & 89 & \textbf{87}\\
	\end{tabular}
\end{table*}
\section{Conclusion}\label{sec:conclusion}
In this paper, we present an unsupervised IDS that captures inter-dependencies in high-level basic features in the traffic data. Our IDS uses a novel Autoencoder combining multi-scale Convolutional Neural Network with Long Short-Term Memory (MSCNN-LSTM) in the Autoencoder architecture to capture spatial-temporal dependencies in traffic data. The performance of the proposed approach was evaluated using the NSL-KDD, UNSW-NB15, and CICDDoS2019 datasets respectively. The experimental results show that the proposed approach not only has good detection performance but also obtains high recall and F-score compare to the existing IDS models in the literature. The proposed approach can be further improved by using different feature selection techniques, optimizing hyper-parameters, and fine-tuning the model's architecture. In our future work, we will extend this model for the multi-class scenarios to detect different classes of intrusion attacks specifically minority attacks. We also plan to apply the proposed method for Android-based malware detection \cite{zhu2020multi, zhu2021task}, or ransomware detection and classification tasks \cite{zhu2021few, mcintosh2018large, mcintosh2019inadequacy} to evaluate the generalizability and practicability.

\section*{Acknowledgment}
This research is supported by the Cyber Security Research Programme—Artificial Intelligence for Automating Response to Threats from the Ministry of Business, Innovation, and Employment (MBIE) of New Zealand as a part of the Catalyst Strategy Funds under the
grant number MAUX1912.

\bibliography{Bibliography}

\end{document}